\documentclass[12pt]{article}
\usepackage{a4wide}
\usepackage{latexsym}
\usepackage{amsmath}
\usepackage{amsfonts}
\usepackage{cite}

\usepackage{pslatex}
\usepackage[latin1]{inputenc}
\usepackage[T1]{fontenc}

\def\bq{\begin{eqnarray}}
\def\eq{\end{eqnarray}}
\def\l{\langle}
\def\r{\rangle} 
\def\eps{\varepsilon}

\begin{document}

\thispagestyle{empty}

\begin{flushright}
  MZ-TH/11-18
\end{flushright}

\vspace{1.5cm}

\begin{center}
  {\Large\bf Does one need the $\cal{O}(\eps)$- and ${\cal O}(\eps^2)$-terms of one-loop amplitudes in an NNLO calculation ?\\
  }
  \vspace{1cm}
  {\large Stefan Weinzierl\\
\vspace{2mm}
      {\small \em Institut f{\"u}r Physik, Universit{\"a}t Mainz,}\\
      {\small \em D - 55099 Mainz, Germany}\\
  } 
\end{center}

\vspace{2cm}

\begin{abstract}\noindent
  {
This article discusses the occurences of one-loop amplitudes within a next-to-next-to-leading order calculation.
In an NNLO calculation the one-loop amplitude enters squared and one would therefore naively expect that the
$\cal{O}(\eps)$- and ${\cal O}(\eps^2)$-terms of the one-loop amplitudes are required.
I show that the calculation of these terms can be avoided if a method is known, which computes the ${\cal O}(\eps^0)$-terms
of the finite remainder function of the two-loop amplitude.
   }
\end{abstract}

\vspace*{\fill}

\newpage

\section{Introduction}
\label{sec:intro}

Precision calculations in high-energy particle physics require the computation of higher order perturbative corrections.
In the area of jet physics one needs in particular fully differential calculations which allow arbitrary infrared-safe experimental cuts.
Several processes have been calculated in the past to next-to-next-to-leading order (NNLO) 
accuracy \cite{Anastasiou:2004qd,Gehrmann-DeRidder:2004tv,GehrmannDeRidder:2007hr,GehrmannDeRidder:2007bj,GehrmannDeRidder:2008ug,GehrmannDeRidder:2009dp,Weinzierl:2006ij,Weinzierl:2006yt,Weinzierl:2008iv,Weinzierl:2009ms,Weinzierl:2009yz,Weinzierl:2010cw,Anastasiou:2002yz,Anastasiou:2003yy,Anastasiou:2003ds,Anastasiou:2004xq,Anastasiou:2005qj,Anastasiou:2007mz,Melnikov:2006di,Anastasiou:2005pn,Catani:2001ic,Catani:2001cr,Grazzini:2008tf,Catani:2009sm,Harlander:2001is,Harlander:2002wh,Harlander:2003ai,Ravindran:2003um,Ravindran:2004mb}.
Most of these calculations have been done with an approach based on Feynman diagrams.

At the level of next-to-leading order (NLO) calculations there has been a significant breakthrough for multi-particle amplitudes in recent years.
This has been mainly due to unitarity or cut-based techniques \cite{Bern:1995cg,Britto:2004nc,Ossola:2006us,Berger:2010zx,Ita:2011wn,Bevilacqua:2009zn,KeithEllis:2009bu,Melia:2010bm,Frederix:2010ne,Hirschi:2011pa,Mastrolia:2010nb}.
In addition, purely numerical approaches based on subtractions for the computation of multi-parton one-loop amplitudes have been 
developed \cite{Nagy:2003qn,Assadsolimani:2009cz,Assadsolimani:2010ka,Becker:2010ng}.
It is a common feature of these approaches that they no longer rely on Feynman diagrams but work on the level of amplitudes instead.
It is a question of current interest if these methods can be extended to NNLO.
This raises immediately a question related to the one-loop amplitudes.
At NNLO the one-loop amplitude enters squared, and since the expansion in the dimensional regularisation parameter $\eps$ starts at order $(-2)$
one would naively expect that up to order $\eps^0$ the 
$\cal{O}(\eps)$- and ${\cal O}(\eps^2)$-terms of the one-loop amplitude are needed for an NNLO calculation.
In fact, several groups (including the author) have calculated $\cal{O}(\eps)$- and ${\cal O}(\eps^2)$-terms of one-loop amplitudes 
in the past \cite{Anastasiou:2000mv,Anastasiou:2001sv,Glover:2001rd,Garland:2001tf,Garland:2002ak,Moch:2002hm,Korner:2008bn,Kniehl:2008fd,Anastasiou:2008vd}.
However, it is by no means obvious how the new approaches for one-loop amplitudes based on unitarity or subtraction
can be extended to include the higher-order terms in the $\eps$-expansion.
For the unitarity approach the complication arises through the required generalisation of the so-called rational terms beyond ${\cal O}(\eps^0)$,
while the numerical method with subtraction is based on evaluating finite integrals in four space-time dimensions, and thus is a priori insensitive to the 
additional $(-2\eps)$-dimensions.

It is therefore appropriate to investigate first what is really needed for an NNLO calculation.
This is the purpose of this article.
In this paper I will trace every appearance of the one-loop amplitude inside an NNLO calculation.
I will show that the $\cal{O}(\eps)$- and ${\cal O}(\eps^2)$-terms of the one-loop amplitude drop out from the final result provided a method is
known, which computes the ${\cal O}(\eps^0)$-terms of the finite remainder function of the two-loop amplitude.
Therefore what is actually needed is just a method to compute the ${\cal O}(\eps^0)$-terms of the finite remainder function of the two-loop amplitude.
In view of the new techniques for the computation of one-loop amplitudes, this is conceptually simpler than the computation
of the two-loop amplitude up to order ${\cal O}(\eps^0)$ and the computation of the one-loop amplitude up to order ${\cal O}(\eps^2)$.

This paper is organised as follows: 
In the next section the problem is exposed.
Section~\ref{sect_review} introduces the notation for this article and reviews the known results, which will be relevant in the following.
The topics covered in this section are the ultraviolet renormalisation of loop amplitudes, the infrared pole structure of loop amplitudes,
the behaviour of tree and loop amplitudes in singular regions of phase space as well as a short introduction into the subtraction method.
Section~\ref{sect_cancel} is the main section of this article.
In this section I derive a theorem, which states that only the tree-level amplitudes, as well as the ${\cal O}(\eps^0)$-terms
of the finite remainder functions ${\cal F}^{(1)}$ and ${\cal F}^{(2)}$ are needed for an NNLO calculation.
Finally, section~\ref{sect_conclusions} contains the conclusions and an outlook.
In an appendix the changes for going from Catani's original definition of the insertion operators and remainder functions to a definition in a minimal scheme
are discussed in detail.

\section{Exposition of the problem}
\label{sect:problem}

The essential point of the argument can already be explained for the case where all infrared singularities occur in the final state.
Therefore I will discuss the case of electron-positron annihilation first and in detail. 
The case of hadronic initial states requires only minor modifications and is explained at the end of section~\ref{sect_cancel}.
The master formula to calculate an observable at an collider with no 
initial-state hadrons (e.g. an electron-positron collider) is given by
\bq
\label{master_formula}
\l {\cal O} \r & = & \frac{1}{2 K(s)}
             \frac{1}{n_{\mathrm{spin}}(1) n_{\mathrm{spin}}(2)}
             \sum\limits_n
             \int d\phi_{n}
             {\cal O}_n\left(p_1,...,p_n,q_1,q_2\right)
             \left| {\cal A}_{n} \right|^2,
\eq
where $q_1$ and $q_2$ are the momenta of the initial-state particles, 
$2K(s)=2s$ is the flux factor and $s=(q_1+q_2)^2$ is the centre-of-mass energy squared.
The factors $1/n_{\mathrm{spin}}(1)$ and $1/n_{\mathrm{spin}}(2)$ correspond
to an averaging over the spins of the initial particles. $d\phi_{n}$ is the invariant phase space measure
for $n$ final state particles and
${\cal O}_n\left(p_1,...,p_n,q_1,q_2\right)$ is the observable, evaluated with a configuration
depending on $n$ final state partons and two initial state particles.
The amplitudes ${\cal A}_{n}$ are calculated perturbatively.
The leading-order contribution to a $n$-jet observable is given by
\bq
 \left| {\cal A}_n \right|^2_{\mathrm{LO}}
 & = & 
             \left. {\cal A}_n^{(0)} \right.^\ast {\cal A}_n^{(0)}.
\eq
At NLO we have the virtual and the real contribution:
\bq
 \left| {\cal A}_n \right|^2_{\mathrm{NLO}}
 & = & 
             \left. {\cal A}_n^{(0)} \right.^\ast {\cal A}_n^{(1)} 
           + \left. {\cal A}_n^{(1)} \right.^\ast {\cal A}_n^{(0)},  
 \nonumber \\
 \left| {\cal A}_{n+1} \right|^2_{\mathrm{NLO}}
 & = &  
          \left. {\cal A}_{n+1}^{(0)} \right.^\ast {\cal A}_{n+1}^{(0)}.
\eq
At NNLO we have the following contributions:
\bq
 \left| {\cal A}_n \right|^2_{\mathrm{NNLO}}
 & = & 
             \left. {\cal A}_n^{(0)} \right.^\ast {\cal A}_n^{(2)} 
           + \left. {\cal A}_n^{(2)} \right.^\ast {\cal A}_n^{(0)}  
           + \left. {\cal A}_n^{(1)} \right.^\ast {\cal A}_n^{(1)},
 \nonumber \\
 \left| {\cal A}_{n+1} \right|^2_{\mathrm{NNLO}}
 & = &  
          \left. {\cal A}_{n+1}^{(0)} \right.^\ast {\cal A}_{n+1}^{(1)} 
        + \left. {\cal A}_{n+1}^{(1)} \right.^\ast {\cal A}_{n+1}^{(0)},
 \nonumber \\ 
 \left| {\cal A}_{n+2} \right|^2_{\mathrm{NNLO}}
 & = &
   \left. {\cal A}_{n+2}^{(0)} \right.^\ast {\cal A}_{n+2}^{(0)}. 
\eq
Here ${\cal A}_n^{(l)}$ denotes an amplitude with $n$ final-state partons and $l$ loops.
We see that in the NNLO contribution the one-loop amplitude with $n$ final-state partons enters squared
and that the one-loop amplitude with $(n+1)$ final-state partons enters interfered with the corresponding tree amplitude.
Usually the loop amplitudes are calculated within dimensional regularisation.
I denote by $\eps=(4-D)/2$ the regularisation parameter.
The loop amplitudes have a Laurent series expansion in the parameter $\eps$. This series starts 
at order $(-2l)$ for an $l$-loop amplitude.
In particular, for an one-loop amplitude one has
\bq
 {\cal A}_n^{(1)} & = &
 \frac{1}{\eps^2} {\cal A}_n^{(1,-2)} + \frac{1}{\eps} {\cal A}_n^{(1,-1)}
 + {\cal A}_n^{(1,0)} + \eps {\cal A}_n^{(1,1)} + \eps^2 {\cal A}_n^{(1,2)} + {\cal O}\left(\eps^3\right).
\eq
The square of the one-loop amplitude up to order ${\cal O}\left(\eps^0\right)$ is then
\bq
\label{square_one_loop}
\lefteqn{
 \left| {\cal A}_n^{(1)} \right|^2
 = } & & 
 \nonumber \\
 & &
 \frac{1}{\eps^4} \left| {\cal A}_n^{(1,-2)} \right|^2
 +
 \frac{2}{\eps^3} \mathrm{Re} \left( \left. {\cal A}_n^{(1,-2)} \right.^\ast {\cal A}_n^{(1,-1)} \right) 
 +
 \frac{1}{\eps^2} \left[ 
                         \left| {\cal A}_n^{(1,-1)} \right|^2
                       + 2 \; \mathrm{Re} \left( \left. {\cal A}_n^{(1,-2)} \right.^\ast {\cal A}_n^{(1,0)} \right) 
                  \right]
 \nonumber \\
 & &
 +
 \frac{2}{\eps} \mathrm{Re} \left( \left. {\cal A}_n^{(1,-2)} \right.^\ast {\cal A}_n^{(1,1)} 
                       + \left. {\cal A}_n^{(1,-1)} \right.^\ast {\cal A}_n^{(1,0)} \right)
 \nonumber \\
 & &
 + \left| {\cal A}_n^{(1,0)} \right|^2
 +
   2 \; \mathrm{Re} \left( \left. {\cal A}_n^{(1,-2)} \right.^\ast {\cal A}_n^{(1,2)} 
                         + \left. {\cal A}_n^{(1,-1)} \right.^\ast {\cal A}_n^{(1,1)} \right)
 + {\cal O}\left(\eps\right).
\eq
One would therefore naively expect that the $\eps$-part ${\cal A}_n^{(1,1)}$ and the $\eps^2$-part ${\cal A}_n^{(1,2)}$
of the one-loop amplitude ${\cal A}_n^{(1)}$ are needed for an NNLO calculation.
In this paper I will investigate if the calculation of these quantities can be avoided.

A related question concerns the all-order $\eps$-behaviour of the one-loop amplitude ${\cal A}_{n+1}^{(1)}$.
At NNLO ${\cal A}_{n+1}^{(1)}$ enters the contribution of the $(n+1)$-particle final state
\bq
  \frac{1}{2 K(s)}
             \frac{1}{n_{\mathrm{spin}}(1) n_{\mathrm{spin}}(2)}
             \int d\phi_{n+1}
             {\cal O}_{n+1}
             \; 2 \; \mathrm{Re} \left( \left. {\cal A}_{n+1}^{(0)} \right.^\ast {\cal A}_{n+1}^{(1)} \right).
\eq
This expression contributes to an $n$-jet observable, whenever the $(n+1)$ final state particles are
classified by the experimental resolution criteria inside the observable ${\cal O}$ as an
$n$-particle configuration.
This happens when one of the particles becomes soft, or when a pair of particles becomes
collinear.
The phase space integral is actually divergent, and the divergence results from the region where
a massless particle is soft or where two massless particles are collinear.
This is the same situation as in an NLO calculation, where the real emission contribution is given as
a phase space integral over $(n+1)$ final state particles, weighted by the matrix element
\bq
 \left| {\cal A}_{n+1}^{(0)} \right|^2.
\eq
In both case one usually applies the subtraction method \cite{Kunszt:1994mc,Frixione:1996ms,Catani:1997vz,Catani:1997pk,Catani:2002hc,Phaf:2001gc} to handle this problem.
Within the subtraction method one subtracts and adds back a suitable chosen approximation term.
The approximation term must have in all singular limits the same behaviour in $D$ dimensions
as the approximated matrix element.
In this way, the difference between the matrix element and the approximation term is integrable 
over the complete phase space.
In addition the approximation term should be simple enough, such that a phase space integration 
for one particle can be performed analytically.
After this integration has been performed, the expression ``lives'' on an $n$-particle phase space
and is combined with the other terms contributing to the integral over the $n$-particle phase space.

In the beginning of the era of NNLO calculation it was feared, that one would need to know
the all-order $\eps$-behaviour of ${\cal A}_{n+1}^{(1)}$ in order to find a subtraction term,
which approximates this amplitude in $D$ dimensions in the singular limits.
Luckily, it was recognised soon that a one-loop amplitude factorises in the singular limits in a universal way
of the following form \cite{Bern:1994zx,Bern:1998sc,Kosower:1999xi,Kosower:1999rx,Bern:1999ry,Catani:2000pi,Kosower:2003cz}
\bq
\label{factorisation_one_loop}
 \lim\limits_{n+1\rightarrow n} {\cal A}_{n+1}^{(1)} & = &
 \mathrm{Sing}^{(0)} \circ {\cal A}_{n}^{(1)} + \mathrm{Sing}^{(1)} \circ {\cal A}_{n}^{(0)}.
\eq
The notation $\mathrm{Sing} \circ {\cal A}_{n}$ indicates that the factorisation occurs at the level of primitive amplitudes.
$\mathrm{Sing}^{(0)}$ and $\mathrm{Sing}^{(1)}$ are universal functions describing the singular limit
(soft or collinear). 
The universality of the singular functions $\mathrm{Sing}^{(0)}$ and $\mathrm{Sing}^{(1)}$ implies that they do not depend
on the complete set of momenta $\{p_1,...,p_n\}$, but only on two or three external momenta. 
In particular they do not depend on the number $n$.
In order to find the approximation terms, only the $D$-dimensional behaviour of the singular functions
$\mathrm{Sing}^{(0)}$ and $\mathrm{Sing}^{(1)}$ need to be known, not the $D$-dimensional behaviour
of the one-loop amplitude ${\cal A}_{n+1}^{(1)}$.
Therefore, for an NNLO calculation one needs to know only the first three terms of the $\eps$-expansion
of the one-loop amplitude ${\cal A}_{n+1}^{(1)}$:
\bq
 {\cal A}_{n+1}^{(1)} & = & 
 \frac{1}{\eps^2} {\cal A}_{n+1}^{(1,-2)}
 +
 \frac{1}{\eps} {\cal A}_{n+1}^{(1,-1)}
 +
 {\cal A}_{n+1}^{(1,0)}
 + {\cal O}\left(\eps\right).
\eq
This fact is well established \cite{Bern:1994zx,Bern:1998sc,Kosower:1999xi,Kosower:1999rx,Bern:1999ry,Catani:2000pi,Kosower:2003cz}.

One observes that in the factorisation formula eq.~(\ref{factorisation_one_loop}) the one-loop
amplitude ${\cal A}_{n}^{(1)}$ with $n$ particles in the final state appears.
In addition, there is a third place where the one-loop amplitude ${\cal A}_{n}^{(1)}$ appears:
The two-loop amplitude ${\cal A}_{n}^{(2)}$ can be written as a part containing all the poles and a finite remainder.
There is a universal formula due to Catani \cite{Catani:1998bh}, which describes the pole part.
In this formula the one-loop amplitude ${\cal A}_{n}^{(1)}$ enters.

In the following I will show that in the combination of the three occurrences of the one-loop amplitude ${\cal A}_{n}^{(1)}$
the $\cal{O}(\eps)$- and ${\cal O}(\eps^2)$-terms drop out.


\section{Notation and review of known results}
\label{sect_review}

\subsection{Ultraviolet renormalisation of loop amplitudes}

The loop amplitudes have explicit divergences. It is common practise to regulate these divergences
by dimensional regularisation.
Within dimensional regularisation one continues the space-time dimension from four to $D=4-2\eps$. 
The origin of these divergences are either related to ultraviolet or to infrared singularities.
The ultraviolet divergences are removed by redefining the parameters of the theory.
In massless QCD it is sufficient to renormalise the strong coupling.
In the $\overline{\mbox{MS}}$ scheme the relation between
the bare coupling $\alpha_0$ and the renormalised coupling $\alpha_s(\mu^2)$ evaluated
at the renormalisation scale $\mu^2$ reads:
\begin{eqnarray}
\alpha_0 & = & \alpha_s S_\eps^{-1} \mu^{2\eps} \left[ 1 
 -\frac{\beta_0}{2\eps} \left( \frac{\alpha_s}{2\pi} \right)
 + \left( \frac{\beta_0^2}{4\eps^2} - \frac{\beta_1}{8\eps} \right) 
   \left( \frac{\alpha_s}{2\pi} \right)^2
 + {\cal O}(\alpha_s^3) \right],
\end{eqnarray}
where 
\begin{eqnarray}
S_\eps & = & \left( 4 \pi \right)^\eps e^{-\eps\gamma_E} \, ,
\end{eqnarray}
is the typical phase-space volume factor in $D =4-2\eps$ dimensions, 
$\gamma_E$ is Euler's constant,
and $\beta_0$ and $\beta_1$ are the first two coefficients of the QCD $\beta$-function:
\begin{eqnarray}
\beta_0 = \frac{11}{3} C_A - \frac{4}{3} T_R N_f,
&\;\;\;&
\beta_1 = \frac{34}{3} C_A^2 - \frac{20}{3} C_A T_R N_f - 4 C_F T_R N_f,
\end{eqnarray}
with the colour factors
\begin{eqnarray}
C_A = N_c, \;\;\; C_F = \frac{N_c^2-1}{2N_c}, \;\;\; T_R = \frac{1}{2}.
\end{eqnarray}
Let the expansion in the strong coupling 
of the unrenormalised amplitude be
\bq
{\cal A}_{n,\mathrm{bare}} & = & 
 \left( 4 \pi \alpha_0 \right)^{\frac{n-2}{2}}
 \left[
        \hat{\cal A}_{n,\mathrm{bare}}^{(0)} + \left( \frac{\alpha_0}{2\pi} \right) \hat{\cal A}_{n,\mathrm{bare}}^{(1)}
                    + \left( \frac{\alpha_0}{2\pi} \right)^2 \hat{\cal A}_{n,\mathrm{bare}}^{(2)}
        + O(\alpha_s^3)
 \right].
\eq
Then, the renormalised amplitude can be expressed as
\bq
{\cal A}_{n,\mathrm{ren}} & = & 
 \left( 4 \pi \alpha_s \right)^{\frac{n-2}{2}}
 \left( S_\eps^{-1} \mu^{2\eps} \right)^{\frac{n-2}{2}}
 \left[
        \hat{\cal A}_{n,\mathrm{ren}}^{(0)} + \left( \frac{\alpha_s}{2\pi} \right) \hat{\cal A}_{n,\mathrm{ren}}^{(1)}
                    + \left( \frac{\alpha_s}{2\pi} \right)^2 \hat{\cal A}_{n,\mathrm{ren}}^{(2)}
        + O(\alpha_s^3)
 \right].
\eq
The relations between
the renormalised and the bare amplitudes are
given by
\bq
\hat{\cal A}_{n,\mathrm{ren}}^{(0)} & = & \hat{\cal A}_{n,\mathrm{bare}}^{(0)},
 \nonumber \\
\hat{\cal A}_{n,\mathrm{ren}}^{(1)} & = & S_\eps^{-1} \mu^{2\eps} \hat{\cal A}_{n,\mathrm{bare}}^{(1)}
                      - (n-2) \frac{\beta_0}{4\eps} \hat{\cal A}_{n,\mathrm{bare}}^{(0)},
 \nonumber \\
\hat{\cal A}_{n,\mathrm{ren}}^{(2)} & = & S_\eps^{-2} \mu^{4\eps} \hat{\cal A}_{n,\mathrm{bare}}^{(2)}
                      - n \frac{\beta_0}{4\eps} S_\eps^{-1} \mu^{2\eps} \hat{\cal A}_{n,\mathrm{bare}}^{(1)}
                      + \frac{(n-2)}{16} 
                        \left( n \frac{\beta_0^2}{2\eps^2} - \frac{\beta_1}{\eps} \right) \hat{\cal A}_{n,\mathrm{bare}}^{(0)}.
\eq
In this paper I work with renormalised amplitudes and I will drop the subscript ``$\mathrm{ren}$'' in the following.

\subsection{Infrared structure of loop amplitudes}
\label{sec:infrared}

The infrared pole structure of loop amplitudes in the dimensional regularisation parameter $\eps$
is well understood \cite{Catani:1998bh,Sterman:2002qn,Mitov:2006xs,Becher:2009cu,Becher:2009qa,Becher:2009kw,Ferroglia:2009ii,Ferroglia:2009ep,Dixon:2008gr,Gardi:2009qi,Dixon:2009ur,Bierenbaum:2011gg}
and can be stated explicitly for one- and two-loop amplitudes.
If we regard a loop amplitude as a vector in colour space, then the infrared poles
of the loop amplitude can be expressed through an operator acting on this vector
in colour space. 
It is therefore convenient to introduce the colour charge operators ${\bf T}_i$.
The action of a colour charge operator ${\bf T}_i$ for a quark, gluon and antiquark in the final state is given by
\bq
\label{colour_charge_operator_final}
\mbox{quark :} & & 
 {\cal A}^\ast\left(  ... q_i ... \right) \left( T_{ij}^a \right) {\cal A}\left(  ... q_j ... \right), \nonumber \\
\mbox{gluon :} & & 
 {\cal A}^\ast\left(  ... g^c ... \right) \left( i f^{cab} \right) {\cal A}\left(  ... g^b ... \right), \nonumber \\
\mbox{antiquark :} & & 
 {\cal A}^\ast\left(  ... \bar{q}_i ... \right) \left( - T_{ji}^a \right) {\cal A}\left(  ... \bar{q}_j ... \right).
\eq
The corresponding formulae for colour charge operators for a quark, gluon or antiquark in the initial state are
\bq
\label{colour_charge_operator_initial}
\mbox{quark :} & & 
 {\cal A}^\ast\left(  ... \bar{q}_i ... \right) \left( - T_{ji}^a \right) {\cal A}\left(  ... \bar{q}_j ... \right), \nonumber \\
\mbox{gluon :} & & 
 {\cal A}^\ast\left(  ... g^c ... \right) \left( i f^{cab} \right) {\cal A}\left(  ... g^b ... \right), \nonumber \\
\mbox{antiquark :} & & 
 {\cal A}^\ast\left(  ... q_i ... \right) \left( T_{ij}^a \right) {\cal A}\left(  ... q_j ... \right). 
\eq
In the amplitude an incoming quark is denoted as an outgoing antiquark and vice versa.
We start with the one-loop amplitude, which can be written as
\bq
\label{poles_one_loop}
{\cal A}^{(1)}_n & = & 
 {\bf I}^{(1)} {\cal A}^{(0)}_n 
 + {\cal F}^{(1)}_{n}.
\eq
Here ${\bf I}^{(1)}$ contains all infrared double and single poles in $1/\eps$ and
${\cal F}^{(1)}_{n}$ is a finite remainder.
The $\eps$-expansion of ${\cal F}^{(1)}_{n}$ starts at order $\eps^0$:
\bq
 {\cal F}^{(1)}_{n} & = & {\cal F}^{(1,0)}_{n} + \eps {\cal F}^{(1,1)}_{n} + \eps^2 {\cal F}^{(1,2)}_{n} + {\cal O}\left(\eps^3\right).
\eq
At two-loops, the corresponding formula reads:
\bq
\label{poles_two_loop}
{\cal A}^{(2)}_n & = & 
 {\bf I}^{(2)} {\cal A}^{(0)}_n 
 + {\bf I}^{(1)} {\cal A}^{(1)}_n 
 + {\cal F}^{(2)}_{n}.
\eq
Again, the $\eps$-expansion of the remainder function starts at order $\eps^0$:
\bq
 {\cal F}^{(2)}_{n} & = & {\cal F}^{(2,0)}_{n} + \eps {\cal F}^{(2,1)}_{n} + \eps^2 {\cal F}^{(2,2)}_{n} + {\cal O}\left(\eps^3\right).
\eq
One observes that in eq.~(\ref{poles_two_loop}) the one-loop amplitude ${\cal A}^{(1)}_n$ occurs in combination
with the one-loop insertion operator ${\bf I}^{(1)}$.
The one-loop insertion operator ${\bf I}^{(1)}$ is given in the massless case by
\bq
\label{formula_insertion}
{\bf I}^{(1)} & = & 
 \frac{\alpha_s}{2\pi}
 \frac{1}{2} \frac{e^{\eps \gamma_E}}{\Gamma(1-\eps)}  
 \sum\limits_{i} \frac{1}{ {\bf T}_i^2} {\cal V}_i(\eps)
 \sum\limits_{j \neq i} {\bf T}_i {\bf T}_j
 \left( \frac{-2 p_i p_j}{\mu^2} \right)^{-\eps},
\eq
where 
\bq
 {\cal V}_i(\eps) & = &
  {\bf T}_i^2 \frac{1}{\eps^2} + \gamma_i \frac{1}{\eps} \, ,
\eq
and the coefficients ${\bf T}_i^2$ and $\gamma_i$ are
\bq
{\bf T}_q^2 = {\bf T}_{\bar{q}}^2 = C_F,
 \;\;\;
{\bf T}_g^2 = C_A, 
 \;\;\;
\gamma_q = \gamma_{\bar{q}} = \frac{3}{2} C_F,
  \;\;\;
\gamma_g = \frac{\beta_0}{2}.
\eq
In general, the colour operators ${\bf T}_i {\bf T}_j$ give rise to colour
correlations. 
The corresponding formula for the two-loop insertion operator ${\bf I}^{(2)}$ is known, but not needed in the present article.
The generalisation to the massive case is also known \cite{Catani:1998bh,Sterman:2002qn,Mitov:2006xs,Becher:2009cu,Becher:2009qa,Becher:2009kw,Ferroglia:2009ii,Ferroglia:2009ep,Dixon:2008gr,Gardi:2009qi,Dixon:2009ur,Bierenbaum:2011gg}.
It should be noted that eq.~(\ref{formula_insertion}) corresponds to the original definition of the insertion operators 
${\bf I}^{(1)}$ and ${\bf I}^{(2)}$ due to Catani.
Within this definition, the insertion operators  contain apart from the pole terms also terms 
of order $\eps^k$ with $k \ge 0$.
Alternatively, it is possible to define the insertion operators in such a way that they contain only the pole terms.
This is discussed in detail in appendix~\ref{appendix_scheme}.
It should be kept in mind that a different definition for the insertion operators will also change the finite remainder functions
${\cal F}^{(1)}$ and ${\cal F}^{(2)}$.

\subsection{Singular behaviour in phase space}

QCD amplitudes become singular in phase space, when the momenta of one or more external particles become degenerate.
In perturbative calculations this phenomen occurs first in next-to-leading order calculations.
Singularities occur if either two particles become collinear or if one particle becomes soft.
The factorisation properties are most conveniently discussed through decomposing QCD amplitudes into
primitive amplitudes.
We may write
\bq
 {\cal A}_n^{(l)} & = &
 \sum\limits_j {\cal C}_{n,j}^{(l)} \; A_{n,j}^{(l)},
\eq
where the coefficients ${\cal C}_{n,j}^{(l)}$ carry all the colour information. The quantities $A_{n,j}^{(l)}$ are called primitive amplitudes.
Primitive amplitudes have
a fixed cyclic ordering of the QCD partons,
a definite routing of the external fermion lines through the diagram
and a definite particle content circulating in the loop.
In the following I will discuss the factorisation properties of a single primitive amplitude and 
I will therefore drop the subscript $j$.
For tree-level amplitudes one has 
in the soft limit the factorisation
\bq
\lim\limits_{p_j \; \mathrm{soft}}
A_{n+1}^{(0)}(...,p_i,p_j,p_k,...) & = & \mbox{Eik}_3^{(0)}(p_i,p_j,p_k) A_{n}^{(0)}(...,p_i,p_k,...),
\eq
where the eikonal factor is given by
\bq
\mbox{Eik}_3^{(0)}(p_i,p_j,p_k)
 & = & 
 \frac{2 p_i \cdot \eps(p_j)}{s_{ij}} - \frac{2 p_k \cdot \eps(p_j)}{s_{jk}}.
\eq
The square of the eikonal factor is given by
\bq
\label{sqreikonal}
\sum\limits_{\lambda_j}
\left. \; \mbox{Eik}_3^{(0)}(p_i,p_j,p_k) \right.^\ast 
       \mbox{Eik}_3^{(0)}(p_i,p_j,p_k) & = & 
 4 \frac{s_{ik}}{s_{ij}s_{jk}}.
\eq
Soft singularities lead to colour correlations among the primitive amplitudes $A_{n}^{(0)}$.
\\
\\
In the collinear limit one has the factorisation
\bq
\label{factcollinearlimit}
\lim\limits_{p_i || p_j}
A_{n+1}^{(0)}(...,p_i,p_j,...) & = & 
 \sum\limits_{\lambda} \; \mbox{Split}_3^{(0)}(p_i,p_j) \; A_{n}^{(0)}(...,p,...).
\eq
where $p_i$ and $p_j$ are the momenta of two adjacent legs and
the sum is over all polarisations.
Squaring the splitting amplitudes one obtains
\bq
\label{PLO}
P_{3,\mathrm{quark}}^{(0)}(\lambda,\lambda') & = &  \sum\limits_{\lambda_i, \lambda_j}
 u(p,\lambda) \left. \; \mbox{Split}_3^{(0)}(p_i,p_j) \right.^\ast \mbox{Split}_3^{(0)}(p_i,p_j) \;\bar{u}(p,\lambda'),
 \nonumber \\
P_{3,\mathrm{gluon}}^{(0)}(\lambda,\lambda') & = &  \sum\limits_{\lambda_i, \lambda_j}
 \left. \eps^\mu(p,\lambda) \right.^\ast \;
                \left. \mbox{Split}_3^{(0)}(p_i,p_j) \right.^\ast \mbox{Split}_3^{(0)}(p_i,p_j) \; \eps^\nu(p,\lambda').
\eq
$P^{(0)}(\lambda,\lambda')$ is a tensor in spin space. Collinear singularities lead therefore to correlations
in spin space.
The spin-averaged splitting functions are obtained by
\bq
 \langle P_{3,\mathrm{quark}}^{(0)} \rangle & = & \frac{1}{2} \sum\limits_\lambda P_{3,\mathrm{quark}}^{(0)}(\lambda,\lambda),
 \nonumber \\
 \langle P_{3,\mathrm{gluon}}^{(0)} \rangle & = & \frac{1}{2(1-\eps)} \sum\limits_\lambda P_{3,\mathrm{gluon}}^{(0)}(\lambda,\lambda).
\eq
The corresponding formulae for the factorisation of one-loop amplitudes in soft and collinear limits are
\cite{Bern:1994zx,Bern:1998sc,Kosower:1999xi,Kosower:1999rx,Bern:1999ry,Catani:2000pi,Kosower:2003cz}
\bq
\lefteqn{
\lim\limits_{j \; \mathrm{soft}}
A_{n+1}^{(1)}(...,p_i,p_j,p_k,...) = } & &
 \nonumber \\
 & & 
 \mbox{Eik}_3^{(0)}(p_i,p_j,p_k) A_{n}^{(1)}(...,p_i,p_k,...)
 +
 \mbox{Eik}_3^{(1)}(p_i,p_j,p_k) A_{n}^{(0)}(...,p_i,p_k,...),
 \nonumber \\
\lefteqn{
\lim\limits_{p_i || p_j}
A_{n+1}^{(1)}(...,p_i,p_j,...) = } & & 
 \nonumber \\
 & &
 \sum\limits_{\lambda} \; \left( \mbox{Split}_3^{(0)}(p_i,p_j) \; A_{n}^{(1)}(...,p,...)
 +
 \mbox{Split}_3^{(1)}(p_i,p_j) \; A_{n}^{(0)}(...,p,...)
 \right).
\eq
The one-loop singular functions $\mbox{Eik}_3^{(1)}$ and $\mbox{Split}_3^{(1)}$ are known \cite{Bern:1994zx,Bern:1998sc,Kosower:1999xi,Kosower:1999rx,Bern:1999ry,Catani:2000pi,Kosower:2003cz}.
In this article we are concerned with the part proportional to $A_{n}^{(1)}$. Here the tree-level singular functions
$\mbox{Eik}_3^{(0)}$ and $\mbox{Split}_3^{(0)}$
occur. Therefore this part is very similar to the NLO case.

\subsection{The subtraction method}

The individual contributions 
to $\l {\cal O} \r_{NLO}$ and $\l {\cal O} \r_{NNLO}$ are in general infrared divergent, only the sum is finite.
However, these contributions live on different phase spaces, which prevents a naive Monte Carlo integration approach.
To render the individual contributions finite, one adds and subtracts suitable chosen terms.
At NLO we have \cite{Kunszt:1994mc,Frixione:1996ms,Catani:1997vz,Catani:1997pk,Catani:2002hc,Phaf:2001gc}
\bq
\l {\cal O} \r_{NLO} & = & 
   \int \left( {\cal O}_{n+1} \; d\sigma_{n+1}^{(0)} - {\cal O}_{n} \circ d\alpha^{(0,1)}_{n} \right)
 + \int \left( {\cal O}_{n} \; d\sigma_{n}^{(1)} + {\cal O}_{n} \circ d\alpha^{(0,1)}_{n} \right).
\eq
The notation ${\cal O}_{n} \circ d\alpha^{(0,1)}_{n}$ is a reminder, that
in general the approximation is a sum of terms
\bq
{\cal O}_{n} \circ d\alpha^{(0,1)}_{n} & = & \sum {\cal O}_{n} \; d\alpha^{(0,1)}_{n}
\eq
and the mapping used to relate the $n+1$ parton configuration to a $n$ parton configuration
differs in general for each summand.
\\
\\
In a similar way, the NNLO contribution is written as \cite{Weinzierl:2003fx,Weinzierl:2003ra,Weinzierl:2009nz,Gehrmann-DeRidder:2003bm,Gehrmann-DeRidder:2005aw,Gehrmann-DeRidder:2005cm,GehrmannDeRidder:2007jk}
\bq
\label{nnlo_subtraction}
\l {\cal O} \r_{NNLO} & = &
 \int \left( {\cal O}_{n+2} \; d\sigma_{n+2}^{(0)} 
             - {\cal O}_{n+1} \circ d\alpha^{(0,1)}_{n+1}
             - {\cal O}_{n} \circ d\alpha^{(0,2)}_{n} 
      \right) \nonumber \\
& &
 + \int \left( {\cal O}_{n+1} \; d\sigma_{n+1}^{(1)} 
               + {\cal O}_{n+1} \circ d\alpha^{(0,1)}_{n+1}
               - {\cal O}_{n} \circ d\alpha^{(1,1)}_{n}
        \right) \nonumber \\
& & 
 + \int \left( {\cal O}_{n} \; d\sigma_n^{(2)} 
               + {\cal O}_{n} \circ d\alpha^{(0,2)}_{n}
               + {\cal O}_{n} \circ d\alpha^{(1,1)}_{n}
        \right).
\eq
$d\alpha^{(0,1)}_{n+1}$ is the NLO subtraction term for $(n+1)$-parton configurations,
$d\alpha^{(0,2)}_{n}$ and $d\alpha^{(1,1)}_{n}$ are generic NNLO subtraction terms.
These terms can be decomposed further into the following form
\bq
\label{decomp_subtr_terms}
 d\alpha^{(0,1)}_{n+1} & = & d\alpha^{\mathrm{single}}_{n+1}, 
 \nonumber \\
 d\alpha^{(0,2)}_{n} & = & d\alpha^{\mathrm{double}}_{n} + d\alpha^{\mathrm{almost}}_{n} + d\alpha^{\mathrm{soft}}_{n} 
                           - d\alpha^{\mathrm{iterated}}_{n},
 \nonumber \\
 d\alpha^{(1,1)}_{n} & = & d\alpha^{\mathrm{loop}}_{n} + d\alpha^{\mathrm{product}}_{n}
                         - d\alpha^{\mathrm{almost}}_{n} - d\alpha^{\mathrm{soft}}_{n} + d\alpha^{\mathrm{iterated}}_{n}.
\eq
The terms $d\alpha^{\mathrm{single}}_{n}$ and $d\alpha^{\mathrm{loop}}_{n}$ are of relevance here.
$d\alpha^{\mathrm{single}}_{n}$ is the NLO subtraction term.
The term $d\alpha^{\mathrm{loop}}$ approximates the singular behaviour of one-loop amplitudes in the soft and collinear limits.
This term can be written as the sum of two contributions
\bq
 d\alpha^{\mathrm{loop}} & = & d\alpha^{\mathrm{loop},a} + d\alpha^{\mathrm{loop},b},
\eq
such that $d\alpha^{\mathrm{loop},a}$ contains the tree-level approximation functions and one-loop amplitudes,
while $d\alpha^{\mathrm{loop},b}$ contains the one-loop approximation functions and tree-level amplitudes.
In formulae we have
\bq
\lim\limits_{n+1 \rightarrow n} d\alpha^{\mathrm{loop},a}
 & = & 
 \mathrm{Sing}^{(0)} \circ d\sigma_n^{(1)},
 \nonumber \\
\lim\limits_{n+1 \rightarrow n} d\alpha^{\mathrm{loop},b}
 & = & 
 \mathrm{Sing}^{(1)} \circ d\sigma_n^{(0)}.
\eq
The one-loop amplitudes $A_n^{(1)}$ appear in $d\alpha^{\mathrm{loop},a}$ and we will focus on this term in the sequel.
\\
\\
We need the integral over the unresolved phase space of 
$d\alpha^{\mathrm{single}}_{n}$ and $d\alpha^{\mathrm{loop},a}_{n}$.
For the NLO subtraction term we have
\bq
\label{unresolved_NLO}
 \int {\cal O}_{n} \circ d\alpha^{\mathrm{single}}_{n}
 & = &
  \frac{1}{2 K(s)}
             \frac{1}{n_{\mathrm{spin}}(1) n_{\mathrm{spin}}(2)}
 \int d\phi_n {\cal O}_{n} \left. {\cal A}_n^{(0)} \right.^\ast \left( {\bf I}_{\mathrm{real}}^{(1)} + {\bf F}_{\mathrm{real}}^{(1)} \right) {\cal A}_n^{(0)},
\eq
where the operator ${\bf I}_{\mathrm{real}}^{(1)}$ contains all the poles in the dimensional regularisation parameter $\eps$.
${\bf I}_{\mathrm{real}}^{(1)}$ is independent of the subtraction scheme.
${\bf F}_{\mathrm{real}}^{(1)}$ is a finite remainder and depends on the subtraction scheme.
${\bf I}_{\mathrm{real}}^{(1)}$ is given in massless QCD by
\bq
\label{insertion_real}
 {\bf I}_{\mathrm{real}}^{(1)}
 & = &
 -
 \frac{\alpha_s}{2\pi} 
 \frac{e^{\eps \gamma_E}}{\Gamma(1-\eps)}
 \sum\limits_{i} \sum\limits_{j \neq i}
 {\bf T}_i {\bf T}_j 
 \left( \frac{1}{\eps^2} + \frac{\gamma_i}{{\bf T}_i^2} \frac{1}{\eps} 
 \right)
 \left( \frac{\left|2p_ip_j\right|}{\mu^2} \right)^{-\eps}.
\eq
Note that ${\bf I}_{\mathrm{real}}^{(1)}$ is very similar to ${\bf I}^{(1)}$.
The insertion operator
${\bf I}_{\mathrm{real}}^{(1)}$ is obtained from ${\bf I}^{(1)}$ by multiplying with a factor $(-2)$ and by the substitution
\bq
 -2 p_i p_j & \rightarrow &  \left|2p_ip_j\right|.
\eq
For the integral over the unresolved phase space of 
$d\alpha^{\mathrm{loop},a}_{n}$ one has
\bq
\label{unresolved_NNLO}
 \int {\cal O}_{n} \circ d\alpha^{\mathrm{loop},a}_{n}
 & = &
  \frac{1}{2 K(s)}
             \frac{1}{n_{\mathrm{spin}}(1) n_{\mathrm{spin}}(2)}
 \int d\phi_n {\cal O}_{n} \; 2 \; \mathrm{Re} \left. {\cal A}_n^{(0)} \right.^\ast \left( {\bf I}_{\mathrm{real}}^{(1)} + {\bf F}_{\mathrm{real}}^{(1)} \right) {\cal A}_n^{(1)}.
\eq


\section{Cancellation of the higher-order terms}
\label{sect_cancel}

We are now in a position to put all pieces together. It is instructive to study the NLO case first.
For the virtual part we have
\bq
 2 \; \mathrm{Re} \left. {\cal A}_n^{(0)} \right.^\ast {\cal A}_n^{(1)}
 & = & 
 2 \; \mathrm{Re} \left. {\cal A}_n^{(0)} \right.^\ast  {\bf I}^{(1)} {\cal A}^{(0)}_n 
 +
 2 \; \mathrm{Re} \left. {\cal A}_n^{(0)} \right.^\ast {\cal F}^{(1)}_{n}
\eq
The integration of the subtraction terms for the real part adds a term
\bq
 \left. {\cal A}_n^{(0)} \right.^\ast \left( {\bf I}_{\mathrm{real}}^{(1)} + {\bf F}_{\mathrm{real}}^{(1)} \right) {\cal A}_n^{(0)}.
\eq
In the sum we have
\bq
\label{nlo_cancelation}
 \mathrm{Re} \left. {\cal A}_n^{(0)} \right.^\ast  \left( 2 {\bf I}^{(1)} + {\bf I}_{\mathrm{real}}^{(1)} \right) {\cal A}^{(0)}_n 
 +
 2 \; \mathrm{Re} \left. {\cal A}_n^{(0)} \right.^\ast {\cal F}^{(1)}_{n}
 +
 \left. {\cal A}_n^{(0)} \right.^\ast {\bf F}_{\mathrm{real}}^{(1)} {\cal A}_n^{(0)}.
\eq
The higher-order terms of the one-loop amplitude are contained in the remainder function ${\cal F}^{(1)}_{n}$.
This function gets interfered with tree amplitude ${\cal A}_n^{(0)}$, therefore only the first term ${\cal F}^{(1,0)}_{n}$
of the $\eps$-expansion is needed.
It is worth noting that the combination $2 {\bf I}^{(1)} + {\bf I}_{\mathrm{real}}^{(1)}$ is in general not free of poles:
A single pole remains, whose coefficient is purely imaginary:
\bq
2 {\bf I}^{(1)} + {\bf I}_{\mathrm{real}}^{(1)}
 & = & 
 \frac{i \pi}{\eps}
 \frac{\alpha_s}{2\pi}
 \sum\limits_{i} 
 \sum\limits_{j \neq i} {\bf T}_i {\bf T}_j \Theta\left( -2 p_i p_j \right)
 + {\cal O}\left(\eps^0\right).
\eq
This pole drops out by taking the real part, therefore eq.~(\ref{nlo_cancelation}) is free of poles.

Let us now move to NNLO. We consider all terms, where the one-loop amplitude ${\cal A}_n^{(1)}$ occurs.
First of all, we have the square of the one-loop amplitude ${\cal A}_n^{(1)}$.
We can write this term as
\bq
 \left. {\cal A}_n^{(1)} \right.^\ast {\cal A}_n^{(1)} 
 & = & 
 \left. {\cal A}_n^{(0)} \right.^\ast \left. {\bf I}^{(1)} \right.^\ast {\bf I}^{(1)} {\cal A}_n^{(0)} 
 +
 2 \; \mathrm{Re} \left. {\cal A}_n^{(0)} \right.^\ast \left. {\bf I}^{(1)} \right.^\ast {\cal F}_n^{(1)} 
 +
\left. {\cal F}_n^{(1)} \right.^\ast {\cal F}_n^{(1)}.
\eq
Secondly, from the integration of $d\alpha^{\mathrm{loop},a}_{n}$ we get
\bq
 2 \; \mathrm{Re} \left. {\cal A}_n^{(0)} \right.^\ast \left( {\bf I}_{\mathrm{real}}^{(1)} + {\bf F}_{\mathrm{real}}^{(1)} \right) {\cal A}_n^{(1)}
 & = & 
 2 \; \mathrm{Re} \left. {\cal A}_n^{(0)} \right.^\ast {\bf I}_{\mathrm{real}}^{(1)} {\bf I}^{(1)} {\cal A}_n^{(0)}
 +
 2 \; \mathrm{Re} \left. {\cal A}_n^{(0)} \right.^\ast {\bf I}_{\mathrm{real}}^{(1)} {\cal F}_n^{(1)}
 \nonumber \\
 & &
 +
 2 \; \mathrm{Re} \left. {\cal A}_n^{(0)} \right.^\ast {\bf F}_{\mathrm{real}}^{(1)} {\bf I}^{(1)} {\cal A}_n^{(0)}
 +
 2 \; \mathrm{Re} \left. {\cal A}_n^{(0)} \right.^\ast {\bf F}_{\mathrm{real}}^{(1)} {\cal F}_n^{(1)}.
\eq
Thirdly, the contribution from the two-loop amplitude can be written as
\bq
 2 \; \mathrm{Re} \left. {\cal A}_n^{(0)} \right.^\ast {\cal A}_n^{(2)} 
 & = &
 2 \; \mathrm{Re} \left. {\cal A}_n^{(0)} \right.^\ast {\bf I}^{(1)} {\bf I}^{(1)} {\cal A}^{(0)}_n
 +
 2 \; \mathrm{Re} \left. {\cal A}_n^{(0)} \right.^\ast {\bf I}^{(1)} {\cal F}^{(1)}_n
 \nonumber \\
 & &
 +
 2 \; \mathrm{Re} \left. {\cal A}_n^{(0)} \right.^\ast {\bf I}^{(2)} {\cal A}^{(0)}_n
 +
 2 \; \mathrm{Re} \left. {\cal A}_n^{(0)} \right.^\ast {\cal F}_n^{(2)}.
\eq
Adding these three pieces together we find
\bq
\label{cancelation_nnlo}
\lefteqn{
 2 \; \mathrm{Re} \left. {\cal A}_n^{(0)} \right.^\ast {\cal A}_n^{(2)} 
 +
 \left. {\cal A}_n^{(1)} \right.^\ast {\cal A}_n^{(1)} 
 +
 2 \; \mathrm{Re} \left. {\cal A}_n^{(0)} \right.^\ast \left( {\bf I}_{\mathrm{real}}^{(1)} + {\bf F}_{\mathrm{real}}^{(1)} \right) {\cal A}_n^{(1)}
 = } & &
 \nonumber \\
 & &
 \mathrm{Re} \left. {\cal A}_n^{(0)} \right.^\ast 
 \left( 
        2 {\bf I}^{(2)} 
        + 2 {\bf I}^{(1)} {\bf I}^{(1)} 
        + \left. {\bf I}^{(1)} \right.^\ast {\bf I}^{(1)}
        + 2 {\bf I}_{\mathrm{real}}^{(1)} {\bf I}^{(1)}  
        + 2 {\bf F}_{\mathrm{real}}^{(1)} {\bf I}^{(1)} 
 \right) {\cal A}^{(0)}_n
\nonumber \\
 & &
 +
 2 \; \mathrm{Re} \left. {\cal A}_n^{(0)} \right.^\ast \left( {\bf I}^{(1)}  + \left. {\bf I}^{(1)} \right.^\ast + {\bf I}_{\mathrm{real}}^{(1)} \right) {\cal F}_n^{(1)} 
 \nonumber \\
 & &
 +
 2 \; \mathrm{Re} \left. {\cal A}_n^{(0)} \right.^\ast {\cal F}_n^{(2)}
 +
 \left. {\cal F}_n^{(1)} \right.^\ast {\cal F}_n^{(1)}
 +
 2 \; \mathrm{Re} \left. {\cal A}_n^{(0)} \right.^\ast {\bf F}_{\mathrm{real}}^{(1)} {\cal F}_n^{(1)}.
\eq
The terms in the first line of the r.h.s of eq.~(\ref{cancelation_nnlo}) involve only the Born amplitude ${\cal A}_n^{(0)}$ and the
insertion operators, but do not involve the one-loop finite remainder function ${\cal F}_n^{(1)}$.
Let us then consider the terms in the third line of the r.h.s of eq.~(\ref{cancelation_nnlo}).
It is clear that for the terms $\left. {\cal F}_n^{(1)} \right.^\ast {\cal F}_n^{(1)}$ and 
$ 2 \; \mathrm{Re} \left. {\cal A}_n^{(0)} \right.^\ast {\bf F}_{\mathrm{real}}^{(1)} {\cal F}_n^{(1)}$ only
the $\eps^0$-terms of ${\cal F}_n^{(1)}$ are needed, since ${\cal F}_n^{(1)}$ is multiplied by terms which do not have any poles in $\eps$.
It remains to examine the terms in the second line of eq.~(\ref{cancelation_nnlo}). These are the most critical ones.
They are given by
\bq
 2 \; \mathrm{Re} \left. {\cal A}_n^{(0)} \right.^\ast \left( {\bf I}^{(1)}  + \left. {\bf I}^{(1)} \right.^\ast + {\bf I}_{\mathrm{real}}^{(1)} \right) {\cal F}_n^{(1)} 
\eq
Although the individual insertion operators start at ${\cal O}(\eps^{-2})$, it turns out that in the 
combination of the insertion operators all poles cancel
\bq
 {\bf I}^{(1)}  + \left. {\bf I}^{(1)} \right.^\ast + {\bf I}_{\mathrm{real}}^{(1)} 
 & = & 
 {\cal O}\left(\eps^0\right),
\eq
and it follows, that again only the $\eps^0$-term ${\cal F}_n^{(1,0)}$ of the one-loop remainder function is needed.
We illustrate the cancellation of the poles for the massless case.
Here we have the explicit expression
\bq
\lefteqn{
 {\bf I}^{(1)}  + \left. {\bf I}^{(1)} \right.^\ast + {\bf I}_{\mathrm{real}}^{(1)} 
 = } & &
 \nonumber \\
 & &  
 \frac{\alpha_s}{2\pi}
 \frac{e^{\eps \gamma_E}}{\Gamma(1-\eps)}  
 \sum\limits_{i} 
 \sum\limits_{j \neq i} {\bf T}_i {\bf T}_j
 \left( \frac{1}{\eps^2} + \frac{\gamma_i}{{\bf T}_i^2} \frac{1}{\eps} \right)
 \left[
 \mathrm{Re} \;
 \left( \frac{-2 p_i p_j}{\mu^2} \right)^{-\eps}
 -
 \left( \frac{\left|2p_ip_j\right|}{\mu^2} \right)^{-\eps}
 \right].
\eq
The expression in the square bracket is of order $\eps^2$, therefore the complete expression starts at order $\eps^0$.

For completeness let us also discuss which terms are needed for the finite remainder functions ${\cal F}_n^{(2)}$ and ${\cal F}_{n+1}^{(1)}$.
The two-loop remainder function ${\cal F}_n^{(2)}$ occurs only in the interference term with the Born amplitude ${\cal A}_n^{(0)}$, therefore
it follows that only the $\eps^0$-term ${\cal F}_n^{(2,0)}$ of the two-loop remainder function is needed.
The one-loop remainder function ${\cal F}_{n+1}^{(1)}$ with $(n+1)$ partons occurs only in the interference term with the Born amplitude ${\cal A}_{n+1}^{(0)}$
and in combination with the subtraction term $d\alpha^{\mathrm{loop}}$. This contribution is evaluated in four dimensions, and
therefore only the $\eps^0$-term ${\cal F}_{n+1}^{(1,0)}$ of the one-loop remainder function is needed.

We may summarise the situation as follows:
\\
{\bf Theorem}:
For an NNLO calculation it is sufficient to know the tree-level amplitudes ${\cal A}_n^{(0)}$, ${\cal A}_{n+1}^{(0)}$ and ${\cal A}_{n+2}^{(0)}$,
the ${\cal O}(\eps^0)$-terms of the one-loop finite remainder function ${\cal F}_n^{(1)}$ and ${\cal F}_{n+1}^{(1)}$, as well as
the ${\cal O}(\eps^0)$-terms of the two-loop finite remainder function ${\cal F}_n^{(2)}$.
The $\cal{O}(\eps)$- and ${\cal O}(\eps^2)$-terms of the one-loop finite remainder function ${\cal F}_n^{(1)}$ drop out from the final result
and are therefore not needed.
\\
\\
This is the main result of this paper.
A few remarks are in order:
\\
\\
1.
Although not explicitly shown in this paper, the statement above can be sharpened: Only the four-dimensional results (or equivalently only
the ${\cal O}(\eps^0)$-terms) of the tree-level amplitudes ${\cal A}_n^{(0)}$, ${\cal A}_{n+1}^{(0)}$ and ${\cal A}_{n+2}^{(0)}$ are required.
This follows from the Kinoshita-Lee-Nauenberg theorem, which ensures that all poles cancel in the end. Therefore only finite
expressions multiply the tree-level amplitudes and it is sufficient to know them in four dimensions.
\\
\\
2.
It should be pointed out that the statement above does not say, that the one- and two-loop amplitudes up to ${\cal O}(\eps^0)$ are sufficient.
What needs to be known are the ${\cal O}(\eps^0)$-terms of the finite remainder function ${\cal F}_n^{(2)}$, and not the 
${\cal O}(\eps^0)$-terms of the two-loop amplitude ${\cal A}_n^{(2)}$.
\\
\\
3.
The statement applies also to processes with partons in the initial state.
This can be seen as follows:
For processes with initial-state partons, the integration over the unresolved phase-space 
corresponding to eq.~(\ref{unresolved_NLO}) and eq.~(\ref{unresolved_NNLO}) will give additional poles proportional to the Altarelli-Parisi
splitting functions. 
These poles are canceled by a collinear counterterm arising from the factorisation into a hard scattering process and 
parton distribution functions.
The collinear counterterm is given by
\bq
\lefteqn{
 \int d\sigma^C = 
 \sum\limits_{f_1,f_2}
 \int\limits_0^1 dx_1
 \int\limits_0^1 dx_2
} & &
 \nonumber \\
 & &
  \frac{1}{2 K(\hat{s})}
             \frac{1}{n_{\mathrm{spin}}(1) n_{\mathrm{spin}}(2) n_{\mathrm{colour}}(1) n_{\mathrm{colour}}(2)}
 \sum\limits_n
             \int d\phi_{n}
             {\cal O}_{n}
             f_{f_1}(x_1) f_{f_2}(x_2)
             \left| {\cal A}_{n} \right|^2.
\eq
The first sum is over all partons species $f_1$ and $f_2$ inside the initial-state hadrons. $f_{f_1}(x_1)$ and $f_{f_2}(x_2)$ are the parton
distribution functions. The quantity $\hat{s}$ is the partonic centre-of-mass energy.
The averaging is now also over the colour degrees of freedom of the initial state partons, given by 
$n_{\mathrm{colour}}(1)$ and $n_{\mathrm{colour}}(2)$.
Expanded to next-to-leading order, the collinear counterterm yields a contribution proportional to
\bq
& &
 - \frac{\alpha_s}{2\pi} S_\eps 
 \int\limits_0^1 dz_1 \int\limits_0^1 dz_2
 \left\{ 
   \delta\left(1-z_2\right) \left[ - \frac{1}{\eps} \left( \frac{\mu_F^2}{\mu^2} \right)^{-\eps} P^{(1)}\left(z_1\right) + K^{(1)}\left(z_1\right) \right] 
 \right. \nonumber \\
 & & \left.
 +
   \delta\left(1-z_1\right) \left[ - \frac{1}{\eps} \left( \frac{\mu_F^2}{\mu^2} \right)^{-\eps} P^{(1)}\left(z_2\right) + K^{(1)}\left(z_2\right) \right] 
 \right\} \left| {\cal A}_{n}^{(0)}\left(p_1,...,p_n,z_1 q_1, z_2 q_2\right) \right|^2.
\eq
$P^{(1)}$ are NLO Altrarelli-Parisi splitting functions. The functions $K^{(1)}$ are finite and define the factorisation scheme.
In the $\overline{\mathrm{MS}}$-scheme one has $K^{(1)}=0$.
At NNLO the part involving the one-loop amplitude ${\cal A}_n^{(1)}$ is very similar:
\bq
& &
 - \frac{\alpha_s}{2\pi} S_\eps 
 \int\limits_0^1 dz_1 \int\limits_0^1 dz_2
 \left\{ 
   \delta\left(1-z_2\right) \left[ - \frac{1}{\eps} \left( \frac{\mu_F^2}{\mu^2} \right)^{-\eps} P^{(1)}\left(z_1\right) + K^{(1)}\left(z_1\right) \right] 
 \right. \nonumber \\
 & & \left.
 +
   \delta\left(1-z_1\right) \left[ - \frac{1}{\eps} \left( \frac{\mu_F^2}{\mu^2} \right)^{-\eps} P^{(1)}\left(z_2\right) + K^{(1)}\left(z_2\right) \right] 
 \right\} 
 \; 2 \; \mathrm{Re} \left. {\cal A}_{n}^{(0)} \right.^\ast {\cal A}_{n}^{(1)},
\eq
where it should be understood that the amplitudes are again evaluated with the rescaled momenta $z_1 q_1$ and $z_2 q_2$ for the initial-state partons.
The combination with the additional poles from the integration over the unresolved phase-space 
corresponding to eq.~(\ref{unresolved_NLO}) and eq.~(\ref{unresolved_NLO}) yields in the NLO case the finite contribution
\bq
 \int\limits_0^1 dz_1 \int\limits_0^1 dz_2
 \left\{ 
   \delta\left(1-z_2\right) \left[ {\bf P}^{(1)}\left(z_1\right) + {\bf K}^{(1)}\left(z_1\right) \right] 
 +
   \delta\left(1-z_1\right) \left[ {\bf P}^{(1)}\left(z_2\right) + {\bf K}^{(1)}\left(z_2\right) \right] 
 \right\} \left| {\cal A}_{n}^{(0)} \right|^2.
\eq
The colour-charge operators ${\bf P}^{(1)}$ and ${\bf K}^{(1)}$ are finite. Their precise definition can be found in ref.~\cite{Catani:1997vz}.
In the same way one obtains at NNLO for the part involving the one-loop amplitude ${\cal A}_n^{(1)}$:
\bq
\lefteqn{
 \int\limits_0^1 dz_1 \int\limits_0^1 dz_2
} & &  \\
 & &
 \left\{ 
   \delta\left(1-z_2\right) \left[ {\bf P}^{(1)}\left(z_1\right) + {\bf K}^{(1)}\left(z_1\right) \right] 
 +
   \delta\left(1-z_1\right) \left[ {\bf P}^{(1)}\left(z_2\right) + {\bf K}^{(1)}\left(z_2\right) \right] 
 \right\} 
  \; 2 \; \mathrm{Re} \left. {\cal A}_{n}^{(0)} \right.^\ast {\cal A}_{n}^{(1)}.
 \nonumber
\eq
Since ${\bf P}^{(1)}$ and ${\bf K}^{(1)}$ are finite, the $\cal{O}(\eps)$- and ${\cal O}(\eps^2)$-terms of the one-loop amplitude
are not relevant for this contribution.
This completes the proof of the theorem for the case of initial-state partons.
\\
\\
At the end I would like to discuss a possible application:
Turning eq.~(\ref{poles_one_loop}) and eq.~(\ref{poles_two_loop}) around, we have
\bq
 {\cal F}^{(1)}_{n}
 & = & 
 {\cal A}^{(1)}_n - {\bf I}^{(1)} {\cal A}^{(0)}_n,
 \nonumber \\
 {\cal F}^{(2)}_{n}
 & = &
 {\cal A}^{(2)}_n 
 - {\bf I}^{(1)} {\cal A}^{(1)}_n 
 - {\bf I}^{(2)} {\cal A}^{(0)}_n.
\eq
The remainder functions ${\cal F}^{(1)}_{n}$ and ${\cal F}^{(2)}_{n}$ are finite quantities. 
The ${\cal O}\left(\eps^0\right)$-terms can be calculated numerically in four dimensions as follows:
One first introduces an integral representation for the insertion operators ${\bf I}^{(1)} $ and ${\bf I}^{(2)}$
in $D$ dimensions.
The right-hand side can then be viewed as a single integral over
\bq
 \int \frac{d^Dk}{(2\pi)^D}
\eq
for ${\cal F}^{(1)}_{n}$ and as a single integral over
\bq
 \int \frac{d^Dk_1}{(2\pi)^D} \frac{d^Dk_2}{(2\pi)^D}
\eq
for ${\cal F}^{(2)}_{n}$.
In the next step one adds an approximation term, which approximates the integrand locally in all singular limits and which integrates to zero.
In the final step the limit $D \rightarrow 4$ can be taken and the integrals can be evaluated in four dimensions.
In particular the integrals can be evaluated numerically by Monte Carlo methods.
At NLO this approach has been studied in \cite{Nagy:2003qn,Assadsolimani:2009cz,Assadsolimani:2010ka,Becker:2010ng}.
The present article shows that this method can be generalised to NNLO, bypassing the need to calculate the
$\cal{O}(\eps)$- and ${\cal O}(\eps^2)$-terms of the one-loop amplitudes.


\section{Conclusions}
\label{sect_conclusions}

In this article I traced the appearance of one-loop amplitudes within an NNLO calculation.
I investigated whether the calculation of the $\cal{O}(\eps)$- and ${\cal O}(\eps^2)$-terms of the one-loop amplitudes 
can be avoided.
I showed that the calculation of these terms can be avoided, if a method is known which allows the 
computation of the ${\cal O}(\eps^0)$-terms of the finite remainder functions for the two-loop amplitude
and the one-loop amplitude.
This opens the door for a numerical calculation at NNLO with the help of the subtraction method.
The result of this paper will also be useful for NNLO calculations within the unitarity or cut-based method.

\subsection*{Acknowledgements}

I would like to thank Simon Pl\"atzer for useful discussions on this subject.


\begin{appendix}

\section{Scheme dependence of the insertion operators}
\label{appendix_scheme}

In the main text of this article I have used the original definition of the one-loop and two-loop insertion operators
${\bf I}^{(1)}$ and ${\bf I}^{(2)}$ due to Catani.
With this definition the insertion operators do contain apart from the pole terms also terms of order $\eps^k$ with $k \ge 0$.
It should be mentioned that these insertion operators can also be defined in a minimal scheme, such that in this scheme they contain
exactly the pole terms and nothing else \cite{Becher:2009cu,Becher:2009qa}.
I will denote the insertion operators in the minimal scheme by ${\bf Z}^{(1)}$ and ${\bf Z}^{(2)}$.
The relation between the two schemes is given by
\bq
 {\bf Z}^{(1)} & = & R\left( {\bf I}^{(1)} \right),
 \nonumber \\
 {\bf Z}^{(2)} & = & R\left( {\bf I}^{(2)} + {\bf I}^{(1)} R \left( {\bf I}^{(1)} \right)\right),
\eq
where $R$ denotes the projection onto the pole part:
\bq
 R\left( \sum\limits_{j=-2l}^\infty c_j \eps^j \right)
 & = & 
 \sum\limits_{j=-2l}^{-1} c_j \eps^j.
\eq
Changing the insertion operators to the minimal scheme will also change the finite remainder functions.
One has
\bq
{\cal A}^{(1)}_n & = & 
 {\bf Z}^{(1)} {\cal A}^{(0)}_n 
 + {\cal F}^{(1)}_{n,\mathrm{minimal}},
 \nonumber \\
{\cal A}^{(2)}_n & = & 
 \left( {\bf Z}^{(2)} - {\bf Z}^{(1)} {\bf Z}^{(1)} \right) {\cal A}^{(0)}_n 
 + {\bf Z}^{(1)} {\cal A}^{(1)}_n 
 + {\cal F}^{(2)}_{n,\mathrm{minimal}}.
\eq
The result of this paper is not affected by this change of schemes. Therefore the knowledge of the ${\cal O}(\eps^0)$-terms of the
finite remainder functions ${\cal F}^{(1)}_{n,\mathrm{minimal}}$ and ${\cal F}^{(2)}_{n,\mathrm{minimal}}$
is equally sufficient for an NNLO calculation.

\end{appendix}


\bibliography{/home/stefanw/notes/biblio}
\bibliographystyle{/home/stefanw/latex-style/h-physrev3}

\end{document}